\documentclass{JHEP3}
\usepackage{amsmath}
\usepackage{amsfonts}
\usepackage{amssymb}
\usepackage{amscd}
\usepackage{graphicx}
\usepackage{citesort}
\usepackage{mathrsfs}
\usepackage{bbm}
\usepackage{amsmath, amssymb, graphics}
\def\S{{\mathbb S}}

\newcommand{\comm}[2]{[#1,#2]}
\newcommand{\gen}[1]{\mathfrak{#1}}
\newcommand{\cwgen}[1]{#1}

\def\[{\begin{equation}}
\def\]{\end{equation}}
\def\<{\begin{eqnarray}}
\def\>{\end{eqnarray}}

\newcommand{\half}{\frac{1}{2}}

\newcommand{\adjb}[2]{\text{ad}_{#1}(#2)}

\newcommand{\acomm}[2]{\{#1,#2\}}

\newcommand{\nn}{\nonumber}
\newcommand{\genY}[1]{\widehat{\mathfrak{#1}}}

\newcommand{\alg}[1]{\mathfrak{#1}}
\newcommand{\su}{\alg{su}}

\newcommand{\mathsym}[1]{{}}

\newcommand{\stateA}[1]{|#1\rangle^{\rm{I}}}
\newcommand{\stateB}[1]{|#1\rangle^{\rm{II}}}
\newcommand{\stateC}[1]{|#1\rangle^{\rm{III}}}

\usepackage{amsmath}
\usepackage{amsfonts}
\usepackage{amssymb}
\usepackage{amscd}
\usepackage{graphicx}
\usepackage{citesort}
\usepackage{mathrsfs}
\usepackage{bbm}

\def\S{{\mathbb S}}

\def\ads{{\rm AdS}_5\times {\rm S}^5}

\author{G. Arutyunov\footnote{ Correspondent fellow at
Steklov Mathematical Institute, Moscow.}, M. de Leeuw, A.
Torrielli \footnote{E-mail:
G.E.Arutyunov,M.deLeeuw,A.Torrielli@uu.nl}
 \\  {\it Institute for Theoretical
Physics and Spinoza Institute,\\ Utrecht University, 3508 TD
Utrecht, The Netherlands}}

\abstract{We identify certain blocks in the S-matrix describing
the scattering of bound states of the $\ads$ superstring that
allow for a representation in terms of universal R-matrices of
Yangian doubles. For these cases, we use the formulas for
Drinfeld's second realization of the Yangian in arbitrary
bound-state representations to obtain the explicit expressions for
the corresponding R-matrices. We then show that these expressions
perfectly match  with the previously obtained S-matrix blocks.}

\title{Universal blocks of the AdS/CFT Scattering Matrix}

\preprint{
          \tiny{ITP-UU-09-10}\\[-.5ex]
          \tiny{SPIN-09-10}\\[-.5ex]
          }

\begin{document}

\section{Introduction}

The study of integrable structures in the AdS/CFT correspondence
\cite{Minahan:2002ve,Kazakov:2004qf,Beisert:2004hm,Arutyunov:2004vx,Staudacher:2004tk,Beisert:2005fw,Beisert:2005tm,Hofman:2006xt,Arutyunov:2006ak,Klose:2006zd,Arutyunov:2009ga}
has now developed to a point where the exact solution of the
planar spectral problem seems to be a concrete possibility
\cite{Janik:2006dc,Beisert:2006ib,Beisert:2006ez,Bajnok:2008bm,Fiamberti:2007rj,Zwiebel:2008gr,Bajnok:2008qj,Arutyunov:2009zu,Gromov:2009tv,Bargheer:2009xy,Bombardelli:2009ns,Gromov:2009bc,Arutyunov:2009ur}.
A recent step in this direction was taken in
\cite{Arutyunov:2009mi}, where the S-matrix describing the
scattering of arbitrary bound states of the theory
\cite{Dorey:2006dq} has been obtained by employing the Yangian
symmetry\footnote{For interesting recent developments connecting
Yangian symmetry and scattering amplitudes, see
\cite{Drummond:2009fd}.}
\cite{Torrielli:2007mc,Beisert:2007ds,deLeeuw:2008dp,Moriyama:2007jt,Matsumoto:2007rh,Beisert:2007ty,Spill:2008tp,Matsumoto:2008ww,deLeeuw:2008ye,Spill:2008yr,Matsumoto:2009rf}
. In turn, the knowledge of the exact scattering matrix for
generic states in the spectrum might be essential for better
understanding the TBA approach \cite{Arutyunov:2007tc}.

\smallskip

The S-matrix found in \cite{Arutyunov:2009mi} has a very
interesting yet complicated structure. Understanding this
structure could reveal a great deal of information on the
underlying theory, by adopting the same ideology as for the case
of the S-matrix for fundamental magnons. One immediate consequence
of the result of \cite{Arutyunov:2009mi} is that the S-matrix for
arbitrary bound states is, by construction, given by a formula of
the type

\begin{eqnarray}
\nonumber R = \Lambda^{op} \Lambda^{-1}
\end{eqnarray}
for a certain (super)matrix $\Lambda$. This naturally realizes the
idea of factorizing (Drinfeld's) twists \cite{twi}, and expresses
the ``triangular" factorization of the S-matrix\footnote{We remark
that, even in the case of the fundamental magnon S-matrix, this
fact had not been explicitly shown before.}.

Given the overall difficulty of the problem, it is natural to
start from reducing the S-matrix to smaller subsectors, and trying
to reconduct them to well-understood mathematical objects. One
expects the complete algebraic structure responsible for
integrability to be some complicated and likely new type of
quantum supergroup, whose properties have so far only appeared as
pieces of a bigger puzzle. We intend to provide here another set
of such pieces, coming from selected subsectors of the bound state
S-matrix.

\smallskip

Our first focus will be on a particular subspace of states, which
was the essential starting point of the construction in
\cite{Arutyunov:2009mi}. This is the space of bound states of the
form (\ref{eqn;BasisCase1}) (Case I). The S-matrix transforms
these states among themselves, with amplitudes controlled by a
hypergeometric function (see (\ref{hypergeom}) below). Since only
the bosonic indices are allowed to be transformed, these states
naturally host a representation of one of the $\alg{su}(2)$ (the
``bosonic" one) inside the (centrally-extended) $\alg{psu}(2|2)$
Yangian, and it is a natural question to ask whether the S-matrix
in this subspace is the representation of the universal R-matrix
of the $\alg{su}(2)$ Yangian double, $DY(\alg{su}(2))$. We will
show that this is indeed the case. This will be done by using the
suitable evaluation representations for Drinfeld's second
realization of $DY(\alg{su}(2))$, originally obtained by
\cite{Khoroshkin:1994uk}, in the operator formalism of
\cite{Arutyunov:2008zt}. We will then obtain explicit expressions
for the corresponding R-matrices in terms of hypergeometric
functions.

\smallskip

We remark that the universal R-matrix and its realization in
concrete representations is a well-studied subject in the
mathematical literature, see {\it e.g.}
\cite{Khoroshkin:1994uk,Kulish:1981gi}. Here we will work out  the
expression for the universal R-matrix evaluated in generic
finite-dimensional representations of $\su(2)$ in our particular
basis, suitable for comparison with the blocks of the bound-state
scattering matrix. After completing the necessary steps, we will
provide satisfactory evidence that these expressions match with
our formula for the S-matrix (\ref{hypergeom}) governing the
scattering of Case I states. This matching ought also to be
expected in view of the connection of the hypergeometric function
in (\ref{hypergeom}), shown to be related to a certain $6j$-symbol
in \cite{Arutyunov:2009mi}, with knot theory and the general theory of
solutions of the Yang-Baxter equation \cite{Chari}.

\smallskip

Moreover, one can notice, by analyzing the result of
\cite{Arutyunov:2009mi}, how ubiquitous the amplitudes
$\mathscr{X}^{k,l}_n$ (\ref{hypergeom}) are in the S-matrix for
both Case II and Case III, since by construction they are derived
from the Case I amplitudes. These factors basically encode how the
tail of bosonic states composing the bound states transform, with
the fermions acting temporarily as spectators. The fact that we
can interpret these fundamental building blocks as coming from the
universal R-matrix of $\alg{su}(2)$ suggests that the latter could
be a genuine factor of the full universal R-matrix.

\smallskip

As a second task, we will focus on subspaces consisting of states
with only one species of bosons and one species of fermions. These
are other (four) subspaces, closed under the action of the
S-matrix, and ``transversal" to the Cases listed in
\cite{Arutyunov:2009mi}: namely, they contain vectors from Case I,
II and III, as we will explain below. These subspaces host
representations of $\alg{gl}(1|1)$, and we will show that the
S-matrix in these blocks can be obtained from the universal
R-matrix of the $\alg{gl}(1|1)$ Yangian double,
$DY(\alg{gl}(1|1))$. We will construct the general bound state
representation for (Drinfeld's second realization of)
$DY(\alg{gl}(1|1))$, and then match with our results from
\cite{Arutyunov:2009mi}. The novelty with respect to the
$\alg{su}(2)$ case is represented by the necessity of making an
unusual choice of the evaluation parameter of the Yangian. This is
an indication that these states, when scattering among themselves,
respect an ``effective" $\alg{gl}(1|1)$ Yangian symmetry, whose
parameters somehow encode the effect of the yet-unknown superior
structure of the full universal R-matrix.

\smallskip

The paper is organized as follows. We will first summarize  the
structure and the main features of the bound state S-matrix
constructed in \cite{Arutyunov:2009mi}. We will also concisely
review some notions concerning Drinfeld's first and second
realization of the Yangian. We will then single out the
$\alg{su}(2)$ and the various $\alg{gl}(1|1)$ subspaces,
respectively. We will summarize the corresponding Yangian
representations for arbitrary bound states, and explicitly
evaluate their universal R-matrices, matching with our previously
obtained results. We will conclude with an appendix, containing
some computational details.

\section{Structure of the bound state S-matrix}

In this section, we report a summary of the results of
\cite{Arutyunov:2009mi}, with the aim of fixing the notation and
as a motivation to investigate the universal structure of the
bound state scattering matrix.

We denote the bound state numbers of the scattering particles as
$\ell_1$ and $\ell_2$, respectively. Because of
${\alg{su}(2)}\times {\alg{su}(2)}$ invariance, when the S-matrix
acts on the bound state representation space it leaves five
different subspaces invariant. Two pairs of them are simply
related to each other, therefore only three non-equivalent cases
are given, which we list here below.

\subsection*{Case I}
The standard basis for this vector space, which we will concisely call $V^{\rm{I}}$, is
\begin{eqnarray}\label{eqn;BasisCase1}
\stateA{k,l}\equiv\underbrace{\theta_{3}w_1^{\ell_1-k-1}w_2^{k}}_{\rm{Space
1}}~\underbrace{\vartheta_{3}v_1^{\ell_2-l-1}v_2^{l}}_{\rm{Space
2}},
\end{eqnarray}
for all $k+l=N$. The range of $k,l$ here and in the cases below is
straightforwardly read off from the definition of the states. For
fixed $N$, this gives in this case $N+1$ different vectors. We get
another copy of Case I if we exchange the index $3$ with $4$ in
the fermionic variable, with the same S-matrix.

\subsection*{Case II}

The standard basis for this space $V^{\rm{II}}$ is
\begin{eqnarray}\label{eqn;BasisCase2}
\stateB{k,l}_1&\equiv& \underbrace{\theta_{3}w_1^{\ell_1-k-1}w_2^{k}}~\underbrace{v_1^{\ell_2-l}v_2^{l}},\nonumber\\
\stateB{k,l}_2&\equiv&\underbrace{w_1^{\ell_1-k}w_2^{k}}~\underbrace{\vartheta_{3}v_1^{\ell_2-l-1}v_2^{l}},\\
\stateB{k,l}_3&\equiv&\underbrace{\theta_{3}w_1^{\ell_1-k-1}w_2^{k}}~\underbrace{\vartheta_{3}\vartheta_{4}v_1^{\ell_2-l-1}v_2^{l-1}},\nonumber\\
\stateB{k,l}_4&\equiv&\underbrace{\theta_{3}\theta_{4}w_1^{\ell_1-k-1}w_2^{k-1}}~\underbrace{\vartheta_{3}v_1^{\ell_2-l-1}v_2^{l}},\nonumber
\end{eqnarray}
where $k+l=N$ as before\footnote{We will from now on, with no
risk of confusion, omit indicating ``Space 1" and ``Space 2" under
the curly brackets.}. It is easily seen that we get in this case $4N+2$ states.
Once again, exchanging $3$ with $4$ in the fermionic variable
gives another copy of Case II, with the same S-matrix.

\subsection*{Case III}

For fixed $N=k+l$, the dimension of this vector space $V^{\rm{III}}$ is $6N$. The
standard basis is
\begin{eqnarray}\label{eqn;BasisCase3}
\stateC{k,l}_1&\equiv&\underbrace{w_1^{\ell_1-k}w_2^{k}}~\underbrace{v_1^{\ell_2-l}v_2^{l}},\nonumber\\
\stateC{k,l}_2&\equiv&\underbrace{w_1^{\ell_1-k}w_2^{k}}~\underbrace{\vartheta_{3}\vartheta_{4}v_1^{\ell_2-l-1}v_2^{l-1}},\nonumber\\
\stateC{k,l}_3&\equiv&\underbrace{\theta_{3}\theta_{4}w_1^{\ell_1-k-1}w_2^{k-1}}~\underbrace{v_1^{\ell_2-l}v_2^{l}},\nonumber\\
\stateC{k,l}_4&\equiv&\underbrace{\theta_{3}\theta_{4}w_1^{\ell_1-k-1}w_2^{k-1}}~\underbrace{\vartheta_{3}\vartheta_{4}v_1^{\ell_2-l-1}v_2^{l-1}},\\
\stateC{k,l}_5&\equiv&\underbrace{\theta_{3}w_1^{\ell_1-k-1}w_2^{k}}~\underbrace{\vartheta_{4}v_1^{\ell_2-l}v_2^{l-1}},\nonumber\\
\stateC{k,l}_6&\equiv&\underbrace{\theta_{4}w_1^{\ell_1-k}w_2^{k-1}}~\underbrace{\vartheta_{3}v_1^{\ell_2-l-1}v_2^{l}}\nonumber.
\end{eqnarray}

The different cases are mapped into one another by use of the
(opposite) coproducts of the (Yangian) symmetry generators.

The S-matrix has the following block-diagonal form:
\begin{eqnarray}
\S=\begin{pmatrix}
  \fbox{\small{$\mathscr{X}$}} & ~ & ~ & ~ & ~ \\
  ~ & \fbox{\LARGE{$\mathscr{Y}$}} & ~ & \mbox{\Huge{$0$}} & ~ \\
  ~ & ~ & \fbox{\Huge{$\mathscr{Z}$}} & ~ & ~ \\
  ~ & \mbox{\Huge{$0$}} & ~ & \fbox{\LARGE{$\mathscr{Y}$}} & ~ \\
  ~ & ~ & ~ & ~ & \fbox{\small{$\mathscr{X}$}}
\end{pmatrix}.
\end{eqnarray}
The outer blocks scatter states from $V^{\rm{I}}$
\begin{eqnarray}
&&\mathscr{X}:V^{\rm{I}}\longrightarrow V^{\rm{I}}\\
&&\stateA{k,l}\mapsto \sum_{m=0}^{k+l}
\mathscr{X}^{k,l}_m\stateA{m,k+l-m},
\end{eqnarray}
where $\mathscr{X}^{k,l}_m$ is given by Eq. (4.11) in
\cite{Arutyunov:2009mi}. We will report its explicit expression in
the next section, formula (\ref{hypergeom}). The blocks $\mathscr{Y}$ describe the
scattering of states from $V^{\rm{II}}$
\begin{eqnarray}
&&\mathscr{Y}:V^{\rm{II}}\longrightarrow V^{\rm{II}}\\
&&\stateB{k,l}_j\mapsto \sum_{m=0}^{k+l}\sum_{j=1}^{4}
\mathscr{Y}^{k,l;j}_{m;i}\stateB{m,k+l-m}_j.
\end{eqnarray}
These S-matrix elements are given in Eq. (5.18) of
\cite{Arutyunov:2009mi}, but we will not need their explicit
expression here. Finally, the middle block deals with the third
case
\begin{eqnarray}
&&\mathscr{Z}:V^{\rm{III}}\longrightarrow V^{\rm{III}}\\
&&\stateC{k,l}_j\mapsto \sum_{m=0}^{k+l}\sum_{j=1}^{6}
\mathscr{Z}^{k,l;j}_{m;i}\stateC{m,k+l-m}_j,
\end{eqnarray}
with $\mathscr{Z}^{k,l;j}_{m;i}$ from Eq. (6.11) in
\cite{Arutyunov:2009mi}. Similarly, these expressions are not
needed for the sake of the present discussion, and we refer to \cite{Arutyunov:2009mi} for their details.\smallskip

We recall that the full $\ads$ string bound state S-matrix is then
obtained by taking two copies of the above S-matrix, and
multiplying the result by the {\it square} of the following phase
factor \cite{Chen:2006gq,Arutyunov:2008zt}:
\begin{eqnarray}\label{eqn;FullPhase}
S_{0}(p_{1},p_{2})&=&\left(\frac{x_{1}^{-}}{x_{1}^{+}}\right)^{\frac{\ell_2}{2}}\left(\frac{x_{2}^{+}}{x_{2}^{-}}\right)^{\frac{\ell_1}{2}}\sigma(x_{1},x_{2})\times\nonumber\\
&&\times\sqrt{G(\ell_2-\ell_1)G(\ell_2+\ell_1)}\prod_{q=1}^{\ell_1-1}G(\ell_2-\ell_1+2q),
\end{eqnarray}
where, in our conventions,
\begin{eqnarray}
\label{fattoreG}
G(Q) = \frac{u_1 - u_2 + \frac{Q}{2}}{u_1 - u_2 - \frac{Q}{2}}.
\end{eqnarray}
Here, $u$ is given in the standard variables by
\begin{eqnarray}\label{eqn;defu}
u&\equiv&\frac{g}{4i}
\left(x^++\frac{1}{x^+}+x^-+\frac{1}{x^-}\right).
\end{eqnarray}

\section{Invariant subspaces}

We will describe here in detail the two type of subspaces of
states we will be focussing our attention on.

\subsection{The $\alg{su}(2)$ subspace}

The first subspace is given by states belonging to Case I in the
above classification. We remind that the $\alg{psu}(2|2)$ algebra
has two (``bosonic" and ``fermionic", according to the indices
they transform) $\alg{su}(2)$ subalgebras,  with generators
$\mathbb{L}^a_b$ and $\mathbb{R}^{\alpha}_{\beta}$, respectively.
The first ones satisfy the following commutation relations:
\begin{eqnarray}
\ [\mathbb{L}_{a}^{\ b},\mathbb{J}_{c}] &=& \delta_{c}^{b}\mathbb{J}_{a}-\frac{1}{2}\delta_{a}^{b}\mathbb{J}_{c},\nonumber\\
\ [\mathbb{L}_{a}^{\ b},\mathbb{J}^{c}] &=& -\delta_{a}^{c}\mathbb{J}^{b}+\frac{1}{2}\delta_{a}^{b}\mathbb{J}^{c}.
\end{eqnarray}
The states from Case I form a natural representation on which the
``bosonic" $\alg{su}(2)$ subalgebra of $\mathbb{L}^a_b$'s acts.
The latter transforms the bosonic variables, and leaves the only
two (equal-type) fermions presents as spectators. Furthermore, the
Case I S-matrix satisfies the Yang-Baxter equation by itself, and
it is of difference form. This means that such S-matrix should
naturally come from the universal R-matrix of the $\alg{su}(2)$
Yangian double \cite{Khoroshkin:1994uk}. The Case I S-matrix is
given by \cite{Arutyunov:2009mi}
\begin{eqnarray}
\label{hypergeom} \mathscr{X}^{k,l}_n &=&(-1)^{k+n} \, \pi
\mathcal{D} \frac{\sin [(k-\ell_1) \pi ] \, \Gamma (l+1)}{\sin
[\ell_1 \pi] \sin [(k +l -\ell_2-n) \pi ] \, \Gamma (l-\ell_2+1)
\Gamma
   (n+1)} \times \nonumber\\
&& \frac{\Gamma
   (n+1-\ell_1) \Gamma
   \left(l+\frac{\ell_1-\ell_2}{2}-n-\delta u
   \right) \Gamma \left(1-\frac{\ell_1+\ell_2}{2}-\delta u \right)}{\Gamma
   \left(k+l-\frac{\ell_1+\ell_2}{2}-\delta u +1\right) \Gamma \left(\frac{\ell_1-\ell_2}{2}- \delta u \right)} \times \\
&& _4\tilde{F}_3
   \left(-k,-n,\delta u
   +1-\frac{\ell_1-\ell_2}{2} ,\frac{\ell_2-\ell_1}{2}-\delta u; 1-\ell_1,\ell_2-k-l,l-n+1;1 \right),\nonumber
\end{eqnarray}
where one has defined $_4\tilde{F}_3 (x,y,z,t;r,v,w;\tau) = {_4F_3}
(x,y,z,t;r,v,w;\tau)/[\Gamma (r) \Gamma (v) \Gamma (w)]$ and
\begin{eqnarray}
\label{D} \mathcal{D}= \frac{x_1^- -x_2^+}{x_1^+
-x_2^-}\frac{e^{i\frac{p_1}{2}}}{e^{i\frac{p_2}{2}}}.
\end{eqnarray}
The quantity $\delta u$ equals $u_1 - u_2$, with $u$ given by
(\ref{eqn;defu}). One can check that this S-matrix satisfies the
YBE, and we will indeed show that this formula coincides with what
one obtains from the Yangian universal R-matrix
\cite{Khoroshkin:1994uk}, with the same evaluation parameter $u$
(\ref{eqn;defu}).

\subsection{The $\alg{su}(1|1)$ subspace}

The other subspace we will consider is obtained by restricting the
bound states to having bosonic and fermionic indices of only one
respective type. For definiteness, we will take the bosonic index
to be $1$ and the fermionic index to be $3$. There are four copies
of this subspace, corresponding to the four different choices of
these indices we can make. The embedding of this subspace in the
full bound state representation is spanned by the vectors
\begin{eqnarray}
\left\{\stateC{0,0}_1,~\stateB{0,0}_1,~\stateB{0,0}_2,~\stateA{0,0}\right\}.
\end{eqnarray}
As one can see, this subspace takes particular states from all
three Cases listed above, yet being closed under the action of the
S-matrix. This means that the S-matrix for this subsector
corresponds to a block-diagonal $4\times4$ matrix. The two Case II
states mix with an S-matrix given by formula (3.15) in
\cite{Arutyunov:2009mi}. The amplitude for the Case III state
present is normalized to $1$, while for the Case I state it is
simply the factor $\cal{D}$ (\ref{D}). Putting this together, one
obtains
\begin{eqnarray}\label{eqn;4x4Smat}
\mathbb{S} = \begin{pmatrix}
 1 ~&~ 0 ~&~ 0 ~&~ 0\\
 0 ~&~ e^{-i\frac{p_2}{2}}\frac{x^{+}_{1}-x^{+}_{2}}{x^{+}_{1}-x^{-}_{2}} ~&~ \frac{\sqrt{\ell_1}\eta(p_{1})}{\sqrt{\ell_2}\eta(p_{2})}\frac{x^{+}_{2}-x^{-}_{2}}{x^{+}_{1}-x^{-}_{2}} ~&~ 0\\
 0 ~&~ \frac{e^{i\frac{p_1}{2}}}{e^{i\frac{p_2}{2}}}\frac{\sqrt{\ell_2}\eta(p_{2})}{\sqrt{\ell_1}\eta(p_{1})}\frac{x^{+}_{1}-x^{-}_{1}}{x^{+}_{1}-x^{-}_{2}} ~&~ e^{i\frac{p_1}{2}}\frac{x^{-}_{1}-x^{-}_{2}}{x^{+}_{1}-x^{-}_{2}} ~&~ 0\\
 0 ~&~ 0 ~&~ 0 ~&~ \frac{x_1^-
-x_2^+}{x_1^+ -x_2^-}\frac{e^{i\frac{p_1}{2}}}{e^{i\frac{p_2}{2}}}
\end{pmatrix}.
\end{eqnarray}
We remark that, taken in the fundamental representation, and
suitably un-twisted in order to eliminate the braiding factors
coming from the nontrivial coproduct
\cite{Gomez:2006va,Plefka:2006ze,Arutyunov:2006yd}, this matrix
coincides with the S-matrix of \cite{Beisert:2005wm}. It is
readily checked that this matrix satisfies the Yang-Baxter
equation by itself, therefore it is natural to ask whether it is
the representation of a known (Yangian) universal R-matrix.

In the remainder of the paper we will discuss the universal
R-matrices for the Yangian doubles associated to $\alg{su}(2)$ and
$\alg{gl}(1|1)$, and show that they coincide with the above
discussed bound state S-matrix blocks \cite{Arutyunov:2009mi}. The
construction relies on Drinfeld's second realization of the
Yangian, which we will review in the next section.

\section{Drinfeld's realizations of the Yangian}\label{sec:yang}

In this section we report, for convenience of the reader, the
defining relations of the Yangian of a simple Lie algebra
$\alg{g}$, in Drinfeld's first and second
realization\footnote{When superalgebras will be involved, all
formulas will be understood in their natural graded generalization
(see for example
\cite{stuko,Heckenberger:2007ry,Gow,Spill:2008tp}). The
non-simplicity of $\alg{gl}(1|1)$ will not be an obstacle, as its
Yangian satisfies a similar set of defining relations.}. For a
thorough treatment of the subject, the reader is referred for
instance to \cite{Etingof,Chari,MacKay:2004tc,Molev}.

The first realization \cite{Drin} is obtained as follows. Let
$\alg{g}$ be a finite dimensional simple Lie algebra generated by
$\gen{J}^A$ with commutation relations
$\comm{\gen{J}^A}{\gen{J}^B} = f^{AB}_C\gen{J}^C$, and equipped
with a non-degenerate invariant bilinear form. The Yangian is the
infinite-dimensional (Hopf) algebra generated by level zero
generators $\gen{J}^a$  and level one generators $\genY{J}^a$ \<
\comm{\gen{J}^A}{\gen{J}^B} = f^{AB}_C\gen{J}^C,\\
\comm{\gen{J}^A}{\genY{J}^B} = f^{AB}_C\genY{J}^C, \> subject to
certain (Serre-relations type of) constraints.

The second realization \cite{Dsecond} is given in terms of
generators $\kappa_{i,m}, \xi^\pm_{i,m}$, $i=1,\dots, \text{rank}
\alg{g}$, $m=0,1,2,\dots$ and relations
\begin{align}
\label{def:drinf2}
&[\kappa_{i,m},\kappa_{j,n}]=0,\quad [\kappa_{i,0},\xi^+_{j,m}]=a_{ij} \,\xi^+_{j,m},\nonumber\\
&[\kappa_{i,0},\xi^-_{j,m}]=- a_{ij} \,\xi^-_{j,m},\quad \comm{\xi^+_{j,m}}{\xi^-_{j,n}}=\delta_{i,j}\, \kappa_{j,n+m},\nonumber\\
&[\kappa_{i,m+1},\xi^+_{j,n}]-[\kappa_{i,m},\xi^+_{j,n+1}] = \frac{1}{2} a_{ij} \{\kappa_{i,m},\xi^+_{j,n}\},\nonumber\\
&[\kappa_{i,m+1},\xi^-_{j,n}]-[\kappa_{i,m},\xi^-_{j,n+1}] = - \frac{1}{2} a_{ij} \{\kappa_{i,m},\xi^-_{j,n}\},\nonumber\\
&\comm{\xi^\pm_{i,m+1}}{\xi^\pm_{j,n}}-\comm{\xi^\pm_{i,m}}{\xi^\pm_{j,n+1}} = \pm\frac{1}{2} a_{ij} \acomm{\xi^\pm_{i,m}}{\xi^\pm_{j,n}},\nn\\
&i\neq j,\, \, \, \, n_{ij}=1+|a_{ij}|,\, \, \, \, \, Sym_{\{k\}} [\xi^\pm_{i,k_1},[\xi^\pm_{i,k_2},\dots [\xi^\pm_{i,k_{n_{ij}}}, \xi^\pm_{j,l}]\dots]]=0.
\end{align}
In these formulas, $a_{ij}$ is the (symmetric) Cartan matrix.

Drinfeld \cite{Dsecond} gave the isomorphism between the two
realizations as follows. Let us define a Chevalley-Serre basis for
$\alg{g}$ as composed of Cartan generators $\gen{H}_i$, and
positive (negative) simple roots $\gen{E}^+_i$ ($\gen{E}^-_i$,
respectively). Also, let us define the corresponding Cartan-Weyl
basis for $\alg{g}$, composed of generators $\gen{H}_i$ and
$\cwgen{E}^\pm_\beta$. One has then
\begin{align}
\label{def:isom}
&\kappa_{i,0}=\gen{H}_i,\quad \xi^+_{i,0}=\gen{E}^+_i,\quad \xi^-_{i,0}=\gen{E}^-_i,\nonumber\\
&\kappa_{i,1}=\hat{\gen{H}}_i-v_i,\quad \xi^+_{i,1}=\hat{\gen{E}}^+_i-w_i,\quad \xi^-_{i,1}=\hat{\gen{E}}^-_i-z_i,
\end{align}
where
\<\label{def:specialel}
v_i &=& \frac{1}{4} \sum_{\beta\in\Delta^+}\left(\alpha_i,\beta\right)(\cwgen{E}_\beta^-\cwgen{E}_\beta^+ +  \cwgen{E}_\beta^+\cwgen{E}_\beta^-) - \half\cwgen{H}_i^2, \\
w_i &=& \frac{1}{4}\sum_{\beta\in\Delta^+}  \left(\cwgen{E}_\beta^-\adjb{\gen{E}_i^+}{\cwgen{E}_\beta^+} + \adjb{\gen{E}_i^+}{\cwgen{E}_\beta^+}\cwgen{E}_\beta^- \right) -  \frac{1}{4}\acomm{\gen{E}_i^+}{\gen{H}_i}, \\
z_i &=& \frac{1}{4}\sum_{\beta\in\Delta^+}
\left(\adjb{\cwgen{E}_\beta^-}{\gen{E}_i^-}\cwgen{E}_\beta^+ +
\cwgen{E}_\beta^+ \adjb{\cwgen{E}_\beta^-}{\gen{E}_i^-} \right) -
\frac{1}{4}\acomm{\gen{E}_i^-}{\gen{H}_i} . \> Here $\Delta^+$
denotes the set of positive root vectors, and the adjoint action
is defined as $\adjb{x}{y} = [x,y]$.

The double of the Yangian admits a universal R-matrix which endows
it with a quasi-triangular structure. Explicit formulas have been
given in \cite{Khoroshkin:1994uk} by making use of Drinfeld's
second realization. In the general case the expressions are rather
complicated, therefore we will not report them here. We will
instead report in what follows the concrete examples relevant to
our subspaces of interest.

\section{Universal R-matrix for $\alg{su}(2)$}

We will now proceed to compute the universal R-matrix for the
$\alg{su}(2)$ block of our bound state S-matrix, following
\cite{Khoroshkin:1994uk}. The derivation is split up into three
parts, corresponding to the factorization
\begin{eqnarray}
R = R_E R_H R_F,
\end{eqnarray}
$R_E$ and $R_F$ being ``root" factors, while $R_H$ is a purely
diagonal ``Cartan" factor. As we mentioned in the above, one works
in Drinfeld's second realization of the Yangian. The map
(\ref{def:isom}) between the first and the second realization
becomes in this case
\begin{align}
\label{def:isom2}
&h_{0}={h},\quad e_{0}={e},\quad f_{0}={f},\nonumber\\
&h_{1}=\hat{{h}}-v,\quad e_{1}=\hat{{e}}-w,\quad f_{1}=\hat{{f}} -z,
\end{align}
where
\begin{eqnarray}
v = \frac{1}{2} (\{ f,e\} - h^2 ), \qquad w = - \frac{1}{4} \{ e,h\}, \qquad z = - \frac{1}{4} \{ f,h\}.
\end{eqnarray}
The first realization is given by
\begin{eqnarray}
&&[h,e]=2e, \qquad \qquad \, \, \, \, \, \, \, \, [h,f]=-2f, \qquad \qquad \, \, \, \, \, \, \, \, \, [e,f]=h, \nonumber\\
&&[\hat{h},e]=[h,\hat{e}]=2e, \qquad [\hat{h},f]=[h,\hat{f}]=-2f, \qquad [\hat{e},f]=[e,\hat{f}]=h,
\end{eqnarray}
and in evaluation representation one has $\hat{h}=u h = u(w_2
\partial_{w_2} - w_1 \partial_{w_1})$, $\hat{e}=u e= u (w_2
\partial_{w_1})$ and $\hat{f}=u f= u (w_1 \partial_{w_2})$. By
applying Drinfeld's map (\ref{def:isom2}) to this evaluation
representation, one first finds the level $0$ and $1$ generators
of the second realization. For generic bound state representations
they depend on second order derivatives. After some manipulations,
one can put the level $1$ generators in a simple form, very
suggestive of the possible generalization at level $n$. This form
reads
\begin{eqnarray}
f_n &=& f (u +\frac{h-1}{2} )^n,\\
e_n &=& e (u +\frac{h+1}{2} )^n,\\
h_n &=& e f_n - f e_n.
\end{eqnarray}
These generators coincide with what obtained in
\cite{Khoroshkin:1994uk} for generic highest-weight
representations of $Y(\alg{su}(2))$.

It is easy to check that these generators satisfy the
correct defining relations obtained by specializing (\ref{def:drinf2}):
\begin{align}
\label{relazionizero}
&[h_{m},h_{n}]=0,\quad \, \, \, \, \, \, \, [e_{m},f_{n}]=\, h_{n+m},\nonumber\\
&[h_{0},e_{m}]= 2\,e_{m},\quad [h_{0},f_{m}]=-  2\,f_{m},\nonumber\\
&[h_{m+1},e_{n}]-[h_{m},e_{n+1}] =   \{h_{m},e_{n}\},\nonumber\\
&[h_{m+1},f_{n}]-[h_{m},f_{n+1}] = -  \{h_{m},f_{n}\},\nonumber\\
&[e_{m+1},e_{n}]-[e_{m},e_{n+1}] =   \{e_{m},e_{n}\},\nonumber\\
&[f_{m+1},f_{n}]-[f_{m},f_{n+1}] = -   \{f_{m},f_{n}\}.
\end{align}
The universal $R$-matrix for the double of the Yangian of
$\alg{sl(2)}$ reads
\begin{eqnarray}
\label{univ} &&{ R}={ R}_E { R}_H { R}_F,
\end{eqnarray}
where
\begin{eqnarray}
&&{ R}_E=\prod_{n\ge 0}^{\rightarrow}\exp(- e_n\otimes f_{-n-1}),  \\
&&{ R}_F=\prod_{n\ge 0}^{\leftarrow}\exp(- f_n\otimes e_{-n-1}),  \\
&&{ R}_H=\prod_{n\ge 0} \exp \left\{ {\rm Res}_{u=v}\left[
\frac{\rm d}{{\rm d}u}({ \log }H^+(u))\otimes { \log
}H^-(v+2n+1)\right]\right\}.
\end{eqnarray}
One has defined
\begin{eqnarray}
\label{eqn;Res}
&&{\rm Res}_{u=v}\left(A(u)\otimes B(v)\right)=\sum_k a_k\otimes b_{-k-1}
\end{eqnarray}
for $A(u)=\sum_k a_k u^{-k-1}$ and $B(u)=\sum_k b_k u^{-k-1}$, and the
so-called Drinfeld's currents are given by
\begin{eqnarray}
\label{curr}
&&E^{\pm}(u)=\pm \sum_{n \ge 0 \atop n<0} e_n u^{-n-1} ~,~~~~~~~~
F^{\pm}(u)=\pm \sum_{n \ge 0 \atop n<0} f_n u^{-n-1} \nonumber \\
&&H^{\pm}(u)=1\pm \sum_{n \ge 0 \atop n<0} h_n u^{-n-1} ~.~~~~
\end{eqnarray}

The arrows on the products indicate the ordering one has to follow
in the multiplication, and are a consequence of the normal
ordering prescription for the root factors in the universal
R-matrix \cite{Khoroshkin:1994uk}. For the generic bound state
representations which we have described above, the ordering will
be essential to get the correct result, and cannot be ignored as
it accidentally happens for the case of the fundamental
representation of $\alg{su}(2)$. We will review the computation of
the three relevant factors of the universal R-matrix in Appendix
A, and report here only the final results in our conventions.

We define the state $\langle A,B\rangle\langle C,D\rangle$ as
made of an $A$ number of $w_1$'s, a $B$ number of $w_2$'s in
the first space, and analogously $C$ and $D$ for $v_1$, $v_2$ in
the second space. We also define
\begin{eqnarray}
\label{cd}
c_i &=& u_1-\frac{A-B+1}{2} -i,\nonumber \\
d_i &=& u_2-\frac{C-D-1}{2} +i,\nonumber \\
\tilde{c}_i &=& u_2-\frac{C-D+1}{2} -i,\nonumber \\
\tilde{d}_i &=& u_1-\frac{A-B-1}{2} +i,
\end{eqnarray}
and
$$
\delta u = u_1 - u_2.
$$
One has for the factor $R_F$
\begin{eqnarray}\label{eqn;RE}
\prod^{\leftarrow}_{n\geq0} \exp[-f_n\otimes e_{-1-n}]|k,l\rangle &=& \sum_{m} A_m (A,B,C,D) \, |k-m,l+m\rangle,
\end{eqnarray}
with
\begin{eqnarray}
\label{A}
A_m (A,B,C,D)= m! {B\choose m}{C\choose m} \prod_{i=0}^{m-1}\frac{1}{c_0-d_0-i-m+1}.
\end{eqnarray}
The Cartan part is then given by
\begin{eqnarray}
\label{eqn;RH}
R_H \langle A,B\rangle && \langle C,D\rangle =
\frac{2^{1-2 \delta u} \, \pi  \, \, \Gamma  \big(\frac{2 \delta u +A+B+C-D+2}{2} \big) \, \, \Gamma  \big(\frac{2 \delta u+B- A+C+D+2}{2} \big)}{\Gamma  (\frac{\delta u- A+B+C-D}{2} ) \Gamma  (\frac{\delta u- A+B+C-D+2}{2} ) \Gamma  (\frac{2\delta u -A-B-C-D}{4}
    )}\times \nonumber\\
&&\times\frac{ \Gamma  \big(\frac{2 \delta u-A+B-C-D}{2} \big) \, \, \, \, \, \Gamma  \big( \frac{2 \delta u -A-B+C-D}{2} \big)}{\Gamma  (\frac{2 \delta u+ A+B-C-D+2}{4} ) \Gamma  (\frac{2\delta u -A-B+C+D +2}{4} ) \Gamma
    (\frac{2 \delta u +A+B+C+D+4}{4} )} \langle A,B\rangle\langle C,D\rangle \nonumber\\
&&\nonumber\\
&&\qquad \, \, \, \, \, \, \,  \equiv {\cal{H}}(A,B,C,D) \, \langle A,B\rangle\langle C,D\rangle.
\end{eqnarray}
Finally, the factor $R_E$ is given by
\begin{eqnarray}\label{eqn;RF}
\prod^{\rightarrow}_{n\geq0} \exp[-e_n\otimes f_{-1-n}]|k,l\rangle &=& \sum_{m} B_m |k+m,l-m\rangle,
\end{eqnarray}
where
\begin{eqnarray}
\label{B}
B_m (A,B,C,D) = m! {A\choose m}{D\choose m} \prod_{i=0}^{m-1}\frac{1}{\tilde{d}_0-\tilde{c}_0-i+m-1}.
\end{eqnarray}

We are now ready to put things together and evaluate the action of
the universal R-matrix of $\alg{su}(2)$ on Case I states. From
formulas (\ref{eqn;RE}), (\ref{eqn;RH}) and (\ref{eqn;RF}), we
obtain
\begin{eqnarray}
R |k,l\rangle &=& \sum_{m=0}^{min(B,C)} \, \sum_{n=0}^{min(A,D)+m} \, B_n (A+m,B-m,C-m,D+m) \\
&&\times \, {\cal{H}} (A+m,B-m,C-m,D+m) \, A_m (A,B,C,D) \, |k-m+n,l+m-n\rangle,\nonumber
\end{eqnarray}
where
\begin{eqnarray}
&&A=\ell_1 - k - 1, \qquad B=k,\nonumber\\
&&C=\ell_2 - l - 1, \qquad D=l,
\end{eqnarray}
and the various factors are given by formulas (\ref{A}),
(\ref{eqn;RH}) and (\ref{B}). It is now easy to convert the above
expression into
\begin{eqnarray}
R |k,l\rangle = \sum_{n=0}^{k+l} R_n \, |n,k+l-n\rangle.
\end{eqnarray}
In order to find the amplitudes $R_n$, we proceed as follows.
Taking into account the presence of binomial factors in the
expressions for $A_m$ and $B_n$, which naturally truncate the sum
when $m,n$ lie outside the correct intervals, we can extend the
summation indices to run from $-\infty$ to $\infty$. In this way,
manipulations of the above sums are easier, and one ends up with
\begin{eqnarray}
\label{Rn}
R_n &=& \sum_{m=-n+k}^{\infty} A_m (\ell_1 - k - 1,k,\ell_2-l-1,l)\, \nonumber\\
&&\times \, {\cal{H}} (\ell_1 - k - 1+m,k-m,\ell_2-l-1-m,l+m)\nonumber\\
&&\times \, B_{n-k+m} (\ell_1 - k - 1+m,k-m,\ell_2-l-1-m,l+m).
\end{eqnarray}

The result of the summation can be obtained by restriction to the
suitable integer values of the parameters of the following
meromorphic function, expressed in terms of hypergeometric
functions:
\begin{eqnarray}
&&{\textstyle R_n = a_1
\left[~ _6F_5(\alpha_1,\alpha_2,\alpha_3,\alpha_4,\alpha_5,\alpha_6;\beta_1,\beta_2,\beta_3,\beta_4,\beta_5;1)+ y\times\right.}  \nonumber\\
&&{\textstyle \left.
_6F_5(\alpha_1-1,\alpha_2-1,\alpha_3-1,\alpha_4-1,\alpha_5-1,\alpha_6-1;\beta_1-1,\beta_2-1,\beta_3-1,\beta_4-1,\beta_5-1;1)\right]},\nonumber
\end{eqnarray}
where
\begin{eqnarray}
\begin{array}{lll}
  \alpha_1 = 2+N-n, & \alpha_2 = 1-n, & \alpha_3 = 1+\ell_1-n, \nonumber\\
  \alpha_4 = 2+N-n-\ell_2,~~ & \alpha_5 = 1+N-2n-\delta u +\frac{\ell_1-\ell_2}{2}, ~~& \alpha_6 = 1+l-n-\delta u
  +\frac{\ell_1-\ell_2}{2}~,\nonumber
\end{array}
\end{eqnarray}
and
\begin{eqnarray}
\begin{array}{lll}
  \beta_1 = \alpha_1-l, & \beta_2 = 2+N-\frac{\ell_1+\ell_2}{2}-n-\delta u, & \nonumber\\
  \beta_3 = 1+\frac{\ell_1-\ell_2}{2}-n-\delta u,~~ & \beta_4 = 2+N-n-\delta u +\frac{\ell_1-\ell_2}{2}, ~~& \beta_5 = 1+\frac{\ell_1+\ell_2}{2}-n-\delta
  u.~~\nonumber
\end{array}
\end{eqnarray}
We have also defined

\begin{eqnarray}
{\textstyle y = \frac{(k-n+1) (2 N-\ell_1-\ell_2-2 n-2\delta u+2)
(2N+\ell_1-\ell_2-2 (n+\delta u-1)) \left((\ell_1-2 (n+\delta
u))^2-\ell_2^2\right)}{16 (N-n+1) (N-\ell_2-n+1) n (n-\ell_1) (2
l+\ell_1-\ell_2-2 (n+\delta u))} },
\end{eqnarray}
and
\begin{eqnarray}
&&{\textstyle a_1 =-\frac{(-1)^{k-n} \pi\sin ((n+1) \pi )(N-n+1)
(N-\ell_2-n+1) (n-\ell_1)  (2 (N-2n-\delta u) + \ell_1-\ell_2) (2
(n-l+ \delta u)-\ell_1+\ell_2) } {(k-n+1) (2(N-n-\delta u +
1)-\ell_1-\ell_2) (2(N-n-\delta u+1) + \ell_1-\ell_2)}}\times\nonumber\\
&&~{\textstyle \frac{ \sin \left(\frac{\pi (2(N-2 n- \delta
u-2)+\ell_1-\ell_2)}{4} \right) \sin ^2\left(\frac{\pi  (2 (N- 2n-
\delta u-1)+\ell_1-\ell_2)}{4}
   \right) \sin \left(\frac{\pi  (2( N-2 n- \delta u)+\ell_1-\ell_2)}{4}
   \right)}
{\sin \left(\frac{\pi (2(n+\delta u+1)-\ell_1+\ell_2)}{2}\right)
\sin \left(\frac{\pi (2(n-N+\delta u)-\ell_1+\ell_2)}{2}\right)
\sin \left(\frac{\pi (2(n-N+\delta u)+\ell_1+\ell_2)}{2} \right)
\sin \left(\frac{\pi (\ell_1+\ell_2-2 (n+\delta u+1))}{2} \right)}
}\times\nonumber\\
&&~{\textstyle \frac{\Gamma (k+1) \Gamma (\ell_2-l) \Gamma (1-n)
\Gamma (N-\ell_2-n+1) \Gamma \left(N+\frac{\ell_1-\ell_2}{2}-2
n-\delta u\right) \Gamma\left(l+\frac{\ell_1-\ell_2}{2}-n-\delta
u\right)} {\Gamma (k-n+1) \Gamma
   \left(N-\frac{\ell_1+\ell_2}{2}-n-\delta u+1\right) \Gamma \left(N+\frac{\ell_1-\ell_2}{2}-n-\delta u+1\right)
\Gamma \left(\frac{ 2\delta u +2 - \ell_1-\ell_2}{4} \right)
\Gamma \left(\frac{2\delta u + 4-\ell_1-\ell_2}{4}\right) }
}\times\nonumber\\
&&~{\textstyle \frac{4^{2-\delta u} \mathcal{D}\sin ((n-N+\ell_2)
\pi )}{\Gamma \left(\frac{\ell_1+\ell_2+2\delta u}{4}\right)\Gamma
\left(\frac{\ell_1+\ell_2+2 \delta u+2}{4}\right)
   \Gamma \left(\frac{\ell_1-\ell_2-2 (n+\delta u-1)}{2} \right)
\Gamma \left(\frac{\ell_1+\ell_2-2 (n+\delta u-1)}{2} \right)}}.
\end{eqnarray}
The quantity $N$ is again here equal to $k+l$.

We have checked that this coincides with the r.h.s. of
(\ref{hypergeom}) for a large selection of choices of the integer
parameters, when taking into account
the proper normalization. We have in fact, with the notations of
\cite{Arutyunov:2009mi},
\begin{eqnarray}
\label{macc} {\cal{D}} \, \frac{\Gamma \left( \frac{1}{4}(2 +
\ell_1 - \ell_2 + 2 \delta u)\right) \, \Gamma\left(\frac{1}{4}(2
+ \ell_2 - \ell_1 + 2 \delta u)\right)}{\Gamma\left( \frac{1}{4}(4
- \ell_1 - \ell_2 + 2 \delta u)\right) \,
\Gamma\left(\frac{1}{4}(\ell_1 + \ell_2 + 2 \delta u)\right)} \,
R_n = \mathscr{X}^{k,l}_n.
\end{eqnarray}
The ratio of gamma functions appearing in the above formula is the
(inverse of the) so-called ``character" of the universal R-matrix
in evaluation representations \cite{Khoroshkin:1994uk}, namely its
action on states of highest-weight $\lambda_i = l_i - 1$.

As a remark, we notice that the formulas for the coproducts of the
generators of $DY(\alg{su}(2))$ (as well as for
$DY(\alg{gl}(1|1))$ which will be studied next) are explicitly
known at arbitrary level $n$ in Drinfeld's second realization. It
is easy to see that the coproducts of the level $n=1$ generators
discussed in this section coincide with the truncation to
$\alg{su}(2)$ of the general expressions obtained in
\cite{Spill:2008tp}. This indicates how this smaller Yangian we
have been discussing here can be embedded in the $\alg{psu}(2|2)$
one.


\section{Universal R-matrix for ${\alg{gl}}(1|1)$}

In this section, we focus on four other subsectors of the entire
bound state representation space, closed under the action of the
S-matrix. We will show that the S-matrix block(s) scattering these
sectors can be obtained from the universal R-matrix of a Yangian
double, in suitable evaluation representations.

Each of these sectors is obtained by considering bound states made
of only one type of boson and one type of fermion. The algebra
transforming the states inside these sectors is an
${\alg{sl}}(1|1)$. As it is known, this type of superalgebras
(with a degenerate Cartan matrix) do not admit a universal
R-matrix, therefore we will introduce an extra Cartan generator
\cite{KT} and study the Yangian of the algebra ${\alg{gl}}(1|1)$
instead\footnote{For the purposes of the universal R-matrix, it
will not make any difference to consider real forms of the
algebras when needed.}. Let us start with the canonical
derivation, and adapt the representation later in order to exactly
match with our S-matrix.

We will follow \cite{Khoroshkin:1994uk,Cai:q-alg9709038}. The super
Yangian double $DY\left(gl(1|1)\right)$ is the Hopf algebra
generated by the elements $e_n$, $f_n$, $h_n$, $k_n$, with $n$ an
integer number, satisfying (Drinfeld's second realization)

\begin{eqnarray}
\label{Lie}
&&[h_m ~,~ h_n]=[h_m ~,~ k_n]=[k_m ~,~k_n]=0, \nonumber \\
&&[k_m ~,~ e_n]=[k_m ~,~ f_n] =0, \nonumber \\
&&[h_0~,~e_n]=-2e_n ~,~ [h_0~,~f_n]=2f_n, \nonumber \\
&&[h_{m+1}~,~e_n]-[h_m~,~e_{n+1}]+\{h_m~,~e_n\}=0, \label{dg}\nonumber\\
&&[h_{m+1}~,~f_n]-[h_m~,~f_{n+1}]-\{h_m~,~f_n\}=0, \nonumber \\
&&\{e_m~,~e_n\}=\{f_m~,~f_n\}=0, \nonumber \\
&&\{e_m~,~ f_n\}=-k_{m+n}.
\end{eqnarray}
Drinfeld's currents are given by
\begin{eqnarray}
&&E^{\pm}(t)=\pm \sum_{n \ge 0 \atop n<0} e_n t^{-n-1} ~,~~~~~~~~
F^{\pm}(t)=\pm \sum_{n \ge 0 \atop n<0} f_n t^{-n-1}, \\
&&H^{\pm}(t)=1\pm \sum_{n \ge 0 \atop n<0} h_n t^{-n-1} ~,~~~~
K^{\pm}(t)=1\pm \sum_{n \ge 0 \atop n<0} k_n t^{-n-1}.
\end{eqnarray}
The universal $R$-matrix reads

\begin{eqnarray}
\label{univ11} &&{\cal R}={\cal R}_+{\cal R}_1{\cal R}_2{\cal
R}_-, \label{rmatrix}
\end{eqnarray}
where
\begin{eqnarray}
&&{\cal R}_+=\prod_{n\ge 0}^{\rightarrow}\exp(- e_n\otimes f_{-n-1}),  \\
&&{\cal R}_-=\prod_{n\ge 0}^{\leftarrow}\exp(f_n\otimes e_{-n-1}),  \\
&&{\cal R}_1=\prod_{n\ge 0} \exp \left\{ {\rm Res}_{t=z}\left[(-1)
\frac{ d}{{ d}t}({ \log }H^+(t))\otimes
{\rm ln}K^-(z+2n+1)\right]\right\}, \\
&&{\cal R}_{2}=\prod_{n\ge 0} \exp \left\{ {\rm Res}_{t=z}\left[(-1)
\frac{ d}{{ d}t}({\log }K^+(t))\otimes
{\rm ln}H^-(z+2n+1)\right]\right\},
\end{eqnarray}
and again
\begin{eqnarray}
&&{\rm Res}_{t=z}\left(A(t)\otimes B(z)\right)=\sum_k a_k\otimes b_{-k-1}
\end{eqnarray}
for $A(t)=\sum_k a_k t^{-k-1}$, $B(z)=\sum_k b_k z^{-k-1}$.

One can show that the following bound state representation, acting
on monomials made of a generic bosonic state $v$ and a generic
fermionic state $\theta$, satisfies all the defining relations of
the second realization (\ref{Lie}):
\begin{eqnarray}
&&e_n = \lambda^n \, a \, \theta \partial_v, \qquad  f_n = \lambda^n \, d \, \, v \partial_\theta, \nonumber\\
&&k_n = -\lambda^n \, a d \, (v \partial_v + \theta \partial_\theta),\qquad h_n = (\lambda+\ell-1)^n (v \partial_v - \theta \partial_\theta).
\end{eqnarray}
As usual, we denote by $\ell$ the number of components of the
bound state. At this stage, $a$ and $d$ are arbitrarily chosen
representation labels, and $\lambda$ is a generic spectral
parameter independent of $a,d$. We will later specify the values
they have to take in order to match with the bound state S-matrix
in these subsectors. Let us start by selecting $w_1$ as our boson
$v$, and $\theta_3$ as our fermion $\theta$. Let us also define a
basis of this first subsector in the following way:
\begin{eqnarray}
\left\{\stateC{0,0}_1,~\stateB{0,0}_1,~\stateB{0,0}_2,~\stateA{0,0}\right\}.
\end{eqnarray}
We first compute
\begin{eqnarray}
\mathcal{R}_-=\prod_{n\geq0}^{\leftarrow} \exp [ f_n\otimes e_{-n-1}]
\end{eqnarray}
in our bound state representation. Because of the fermionic nature
of the operators $f_n\otimes e_{-n-1}$, the above expression
simplifies to
\begin{eqnarray}
\mathcal{R}_-&=&1+\sum_{n\geq0}f_n\otimes e_{-n-1}\nonumber\\
&=&1+\sum_{n\geq0} \frac{u_1^n}{u_2^{n+1}}f\otimes e\nonumber\\
&=&1-\frac{f\otimes e}{\delta \lambda}
\end{eqnarray}
Considering that this term will act non-trivially only on a state
with a fermion in the first space\footnote{Tensor products of
generators act according to the rule $(X\otimes Y)|a\rangle\otimes
|b\rangle = (-)^{[Y][a]} X|a\rangle\otimes Y|b\rangle$, where
$[x]$ denotes the fermionic grading of $x$.}, we easily obtain
\begin{eqnarray}
\mathcal{R}_-=
\begin{pmatrix}
  1 & 0 & 0 & 0 \\
  0 & 1 & 0 & 0 \\
  0 & \frac{a_2d_1\ell_2}{\delta \lambda} & 1 & 0 \\
  0 & 0 & 0 & 1
\end{pmatrix}.
\end{eqnarray}
We have defined
\begin{eqnarray}
\delta \lambda = \lambda_1 - \lambda_2.
\end{eqnarray}
Similarly, one finds
\begin{eqnarray}
\mathcal{R}_+&=&1-\sum_{n\geq0}e_n\otimes f_{-n-1}\nonumber\\
&=&1+\frac{e\otimes f}{\delta \lambda},
\end{eqnarray}
which, written in matrix form, looks like
\begin{eqnarray}
\mathcal{R}_+=
\begin{pmatrix}
  1 & 0 & 0 & 0 \\
  0 & 1 & \frac{a_1d_2\ell_1}{\delta \lambda} & 0 \\
  0 & 0 & 1 & 0 \\
  0 & 0 & 0 & 1
\end{pmatrix}.
\end{eqnarray}
Let us now turn to the Cartan part. For this, we first need to
compute the currents. They are found to be
\begin{eqnarray}
H^\pm &=& 1+\frac{h}{1+\lambda-\ell-t},\\
K^\pm &=& 1+\frac{k}{\lambda-t},
\end{eqnarray}
where we used the fact that both $h$ and $k$ are diagonal
operators. In appropriate domains one then has in particular
\begin{eqnarray}
-\frac{d}{dt}\log H^+ = \sum_{m=1}^{\infty}
\left\{{(\lambda+\ell-1)^m} -{(\lambda+\ell-1-h)^m}\right\}
t^{-m-1}
\end{eqnarray}
and
\begin{eqnarray}
\log K^-(z+2n+1) &=& \log K^-(2n+1) + \\
&& +\sum_{m=1}^{\infty} \left\{\frac{1}{(\lambda-1-2n)^m}
-\frac{1}{(\lambda-1-2n-k)^m}\right\} \frac{z^{m}}{m}\nonumber.
\end{eqnarray}
Straightforwardly computing the residue and performing the sum
yields, in matrix form,
\begin{eqnarray}
\mathcal{R}_1 =
\frac{\Gamma \left(\frac{\delta \lambda + \ell_1}{2} \right) \Gamma \left(\frac{\delta \lambda - a_2d_2\ell_2}{2}\right)}{\Gamma \left(\frac{\delta \lambda}{2}\right) \Gamma \left(\frac{\delta \lambda + \ell_1-a_2d_2\ell_2}{2}\right)}
\begin{pmatrix}
  1 & 0 & 0 & 0 \\
  0 & \frac{\delta \lambda-a_2 d_2 \ell_2}{\delta \lambda} & 0 & 0 \\
  0 & 0 & 1 & 0 \\
  0 & 0 & 0 &  \frac{\delta \lambda-a_2 d_2 \ell_2}{\delta \lambda}
\end{pmatrix}.
\end{eqnarray}
One can perform an analogous derivation for $R_2$ and find
\begin{eqnarray}
\mathcal{R}_2 =
\frac{\Gamma \left(\frac{\delta \lambda + a_1d_1\ell_1+2}{2}\right) \Gamma \left(\frac{\delta \lambda-\ell_2+2}{2}\right)}{\Gamma \left(\frac{\delta \lambda+2}{2}\right) \Gamma \left(\frac{\delta \lambda + a_1d_1\ell_1-\ell_2+2}{2}\right)}
\begin{pmatrix}
  1 & 0 & 0 & 0 \\
  0 & 1 & 0 & 0 \\
  0 & 0 & \frac{\delta \lambda}{\delta \lambda + a_1d_1 \ell_1} & 0 \\
  0 & 0 & 0 &  \frac{\delta \lambda}{\delta \lambda + a_1 d_1 \ell_1}
\end{pmatrix}.
\end{eqnarray}
Multiplying everything out finally gives us the universal R-matrix
in our bound state representation:
\begin{eqnarray}
\label{uni}
\mathcal{R} = A
\left(
\begin{array}{cccc}
 1 & 0 & 0 & 0 \\
 0 & 1-\frac{a_2 d_2 \ell_2}{\delta \lambda+a_1 d_1 \ell_1} & \frac{a_1 d_2 \ell_1}{\delta \lambda+a_1
   d_1 \ell_1} & 0 \\
 0 & \frac{a_2 d_1 \ell_2}{\delta \lambda+a_1 d_1 \ell_1} & \frac{\delta \lambda}{\delta \lambda+a_1 d_1 \ell_1} &
   0 \\
 0 & 0 & 0 & \frac{\delta \lambda-a_2 d_2 \ell_2}{\delta \lambda+a_1 d_1 \ell_1}
\end{array}
\right),
\end{eqnarray}
where
\begin{eqnarray}
\label{normA} A=\frac{\Gamma \left(\frac{\delta
\lambda+\ell_1}{2}\right) \Gamma \left(\frac{\delta \lambda+a_1 d_1
   \ell_1+2}{2}\right) \Gamma \left(\frac{\delta \lambda-\ell_2+2}{2}\right) \Gamma \left(\frac{\delta \lambda-a_2
   d_2 \ell_2}{2} \right)}{\Gamma \left(\frac{\delta \lambda}{2}\right) \Gamma \left(\frac{\delta \lambda+2}{2}\right) \Gamma \left(\frac{\delta \lambda+a_1 d_1 \ell_1-\ell_2+2}{2} \right) \Gamma \left(\frac{\delta \lambda+\ell_1-a_2 d_2\ell_2}{2}\right)}.
\end{eqnarray}
For $a_i=d_i=\ell_i=1$ this reduces to the formula in \cite{Cai:q-alg9709038},
\begin{eqnarray}
\mathcal{R} \propto \mathbbm{1} + \frac{P}{\delta \lambda},
\end{eqnarray}
where $P$ is the graded permutation matrix.

But we can also take $a,d$ to be the representation labels of the
supercharges in the centrally extended ${\alg{psu}}(2|2)$
superalgebra, i.e.
\begin{eqnarray}
a = \sqrt{\frac{g}{2\ell}}\eta, \qquad
d=\sqrt{\frac{g}{2\ell}}\frac{x^{+}-x^-}{i\eta}.
\end{eqnarray}
This corresponds to considering the generators $e,f$ as the
restriction to this subsector of the two supercharges
$\mathbb{Q}_{1}^{3}$ and $\mathbb{G}_{3}^{1}$. It is now readily
seen that by choosing $\lambda$ to be $\frac{g}{2i} x^-$, we can
exactly reproduce\footnote{This is similar to the observation in
\cite{Beisert:2005wm} for the the case of the fundamental
representation.} the $4\times4$ block (\ref{eqn;4x4Smat}) from
(\ref{uni}), after we properly normalize it and introduce the
appropriate braiding factors. To normalize, we simply divide the
formula coming from the universal R-matrix by $A$ (\ref{normA}).
To introduce the braiding factors, we need to twist it by
\cite{Arutyunov:2008zt}
$$
U_2^{-1} (p_1) \, \mathcal{R} \, U_1 (p_2),
$$
with $U(p) = diag(1,e^{- i p/2})$.

There is also another choice for $a,d$ from the ${\alg{psu}}(2|2)$
algebra. Namely, one can also restrict the supercharges
$\mathbb{Q}_{2}^{4}$ and $\mathbb{G}_{4}^{2}$ to this sector. This
means that our parameters $a,d$ will now become the $c,b$ from the
bound state representation
\begin{eqnarray}
\begin{array}{lll}
b = \sqrt{\frac{g}{2\ell}}
\frac{i\zeta}{\eta}\left(\frac{x^{+}}{x^{-}}-1\right),& &
  c = -\sqrt{\frac{g}{2\ell}}\frac{\eta}{\zeta x^{+}}.
\end{array}
\end{eqnarray}
Remarkably, in order to match with (\ref{eqn;4x4Smat}), one has to
choose $\lambda = \frac{i g}{2 x^-}$ and $ \zeta_1=\zeta_2$. The
correct braiding factors can be incorporated by means of the
inverse of the above mentioned twist \cite{Arutyunov:2008zt}.

A similar argument can finally be seen to hold for all the other
subsectors corresponding to different fixed bosonic and fermionic
indices.

While it is likely that in the full universal R-matrix (where one
is supposed to have at once all generators of ${\alg{psu}}(2|2)$)
some kind of ``average" of the two situations will
occur\footnote{In the fundamental representation, this is
exemplified by some of the formulas in \cite{Torrielli:2008wi}.},
we have shown here that the S-matrix in these subspaces can be
``effectively" described by the universal R-matrix of
$DY(\alg{gl}(1|1))$ taken in (two inequivalent choices of)
evaluation representations.


\section*{Acknowledgments}
One of us (A.T.) would like to thank J. Plefka and F. Spill for an
early time collaboration on the ${\alg{gl}}(1|1)$ sector. The work of G.~A. was
supported in part by the RFBI grant 08-01-00281-a, by the grant
NSh-672.2006.1, by NWO grant 047017015 and by the INTAS contract
03-51-6346.

\appendix

\section{Revisiting the $Y(\alg{su}(2))$ computation}

In this Appendix, we give the computational details for the
$\alg{su}(2)$ case.

\subsection{The Factor $R_F$}

Let us first compute how $R_F$ acts on an arbitrary Case I state.
We find
\begin{eqnarray}\label{eqn;RE2}
\prod^{\leftarrow}_{n\geq0} \exp[-f_n\otimes e_{-1-n}]|k,l\rangle &=& \sum_{m} A_m |k-m,l+m\rangle.
\end{eqnarray}
The term $A_m$ is built up out of $m$ copies of $-f\otimes e$
acting on the state $\langle A,B\rangle\langle C,D\rangle$, which
is made of an $A$ number of $w_1$'s, a $B$ number of $w_2$'s in
the first space, and analogously $C$ and $D$ for $v_1$, $v_2$ in
the second space. In view of (\ref{eqn;RE2}), we find that such
terms can come from different exponentials, i.e. with different
$n$'s, or from the same exponential. One first needs to know how the
product of $m$ $f$'s acts on the state $\langle A,B\rangle$.

We conveniently define as in the main text
\begin{eqnarray}
c_i &=& u_1-\frac{A-B+1}{2} -i,\\
d_i &=& u_2-\frac{C-D-1}{2} +i.
\end{eqnarray}
In general one has
\begin{eqnarray}
f_{n_m}\ldots f_{n_2} f_{n_1}\langle A,B\rangle &=&f_{n_m}\ldots f\left(u+\frac{h-1}{2}\right)^{n_2}f\left(u+\frac{h-1}{2}\right)^{n_1}\langle A,B\rangle \nonumber\\
&=& f_{n_m}\ldots f\left(u+\frac{h-1}{2}\right)^{n_2} f \left(c_0\right)^{n_1}\langle A,B\rangle \nonumber\\
&=& B\left(c_0\right)^{n_1}f_{n_m}\ldots f\left(u+\frac{h-1}{2}\right)^{n_2}\langle A+1,B-1\rangle \nonumber\\
&=& B(B-1)\left(c_0\right)^{n_1} \left(c_1\right)^{n_2}f_{n_m}\ldots f_{n_3}\langle A+2,B-2\rangle \nonumber\\
&=& \frac{B!}{(B-m)!} c_0^{n_1}\ldots c_{m-1}^{n_m}\langle A+m,B-m\rangle.
\end{eqnarray}
Similar expressions hold for $e_n$ acting on $\langle C,D\rangle$,
but with $d_i$ instead of $c_i$, and producing the state $\langle
C-m,D+m\rangle$. When we consider terms like this coming from the
ordered exponential (\ref{eqn;RE2}), we always have that $n_i\geq
n_{i-1}$. In case $n_i = n_{i+1}$, we also pick up a combinatorial
factor coming from the series of the exponential. Putting all of
this together, we find
\begin{eqnarray}
&&A_m = (-)^m \frac{B!}{(B-m)!}\frac{C!}{(C-m)!}\left\{ \sum_{n_1\leq\ldots\leq n_m} \frac{1}{N(\{n_1,\ldots,n_m\})}\frac{c_0^{n_1}}{d_0^{n_1+1}}\ldots \frac{c_{m-1}^{n_m}}{d_{m-1}^{n_m+1}} \right\}, \nonumber\\
&&N(\{n_1,\ldots,n_m\})=\frac{1}{{\rm ord} S(\{n_1,\ldots,n_m\})}.
\end{eqnarray}
$N$ is a combinatorial factor which is defined as the inverse of
the order of the permutation group of the set
$\{n_1,\ldots,n_m\}$. For example, $N(\{1,1,2\})=\frac{1}{2}$ and
$N(\{1,1,1,2,3,3,4,5\})=\frac{1}{3!}\frac{1}{2!}=\frac{1}{12}$. By
using the fact that $c_i= c_{i+1}+1, d_i= d_{i+1}-1$, one can
evaluate this sum explicitly and find
\begin{eqnarray}
A_m (A,B,C,D)= m! {B\choose m}{C\choose m} \prod_{i=0}^{m-1}\frac{1}{c_0-d_0-i-m+1},
\end{eqnarray}
where we have indicated the dependence on the
parameters $A,B,C,D$ of the state we are acting on. As one can
easily see using (\ref{cd}), the resulting expression is
manifestly of difference form.

\subsection{The Factor $R_H$}

Next is the Cartan part. First, we work out
\begin{eqnarray}
h_n\langle A,B\rangle = \left\{(A+1)B\left[u-\frac{A-B+1}{2}\right]^n - (B+1)A\left[u-\frac{A-B-1}{2}\right]^n \right\}\langle A,B\rangle \nonumber.
\end{eqnarray}
We then recall the definition of $H_\pm$ from (\ref{curr}). From
the explicit realization we give in the main text it follows that
\begin{eqnarray}
H_+(t)\langle A,B\rangle=H_-(t)\langle A,B\rangle =\left\{1-\frac{(A+1)B}{u-t-\frac{1}{2}(A - B +1)}+\frac{A(B+1)}{u-t-\frac{1}{2}(AB-1)}\right\}\langle A,B\rangle.\nonumber
\end{eqnarray}
Defining $K_{\pm} = \log H_{\pm}$, the Cartan part of the
universal R-matrix can be written as
\begin{eqnarray}
R_H = \prod_{n\geq0} \exp\left[{\rm Res}_{t=x}\left(\frac{d}{dt}K_+(t)\otimes K_-(x+2n+1)\right) \right],
\end{eqnarray}
where the residue is defined in (\ref{eqn;Res}). We
have to find the suitable series representations corresponding to
$\frac{d}{dt}K_+(t)$ and $K_-(x+2n+1)$. With an appropriate choice of domains for the variables $t$ and $x$, one can write in particular
\begin{eqnarray}
\frac{d}{dt}K_+(t) &=& \sum_{m\geq 1}\left\{ \alpha_{1}^{m} + \alpha_{2}^{m} - \alpha_{3}^{m} - \alpha_{4}^{m}\right\}t^{-m-1},\\
K_-(x+2n+1) &=& K_-(0) + \sum_{m\geq 1}\left\{ \beta_{1}^{-m} + \beta_{2}^{-m} - \beta_{3}^{-m} - \beta_{4}^{-m}\right\}\frac{x^{m}}{m},
\end{eqnarray}
where
\begin{eqnarray}
\begin{array}{lcl}
    \alpha_1 = u_1+ \frac{1}{2}(A+B+1), & ~ & \alpha_2 = u_1- \frac{1}{2}(A+B+1),\\
    \alpha_3 = u_1- \frac{1}{2}(A-B+1), & ~ & \alpha_4 = u_1- \frac{1}{2}(A-B-1),
\end{array}
\end{eqnarray}
and
\begin{eqnarray}
\begin{array}{lcl}
    \beta_1 = u_2-2n + \frac{1}{2}(D-C-1), & ~ & \beta_2 = u_2-2n+ \frac{1}{2}(D-C-3),\\
    \beta_3 = u_2-2n+ \frac{1}{2}(D+C-1), & ~ & \beta_4 = u_2-2n- \frac{1}{2}(D+C+3),
\end{array}\end{eqnarray}
This leads to
\begin{eqnarray}
\label{eqn;RH2} R_H \langle A,B\rangle && \langle C,D\rangle =
\frac{2^{1-2 \delta u} \, \pi  \, \, \Gamma  \big(\frac{2 \delta u
+A+B+C-D+2}{2} \big) \, \, \Gamma  \big(\frac{2 \delta u+B-
A+C+D+2}{2} \big)}{\Gamma  (\frac{\delta u- A+B+C-D}{2} ) \Gamma
(\frac{\delta u- A+B+C-D+2}{2} ) \Gamma  (\frac{2\delta u
-A-B-C-D}{4}
    )}\times \nonumber\\
&&\times\frac{ \Gamma  \big(\frac{2 \delta u-A+B-C-D}{2} \big) \, \, \, \, \, \Gamma  \big( \frac{2 \delta u -A-B+C-D}{2} \big)}{\Gamma  (\frac{2 \delta u+ A+B-C-D+2}{4} ) \Gamma  (\frac{2\delta u -A-B+C+D +2}{4} ) \Gamma
    (\frac{2 \delta u +A+B+C+D+4}{4} )} \langle A,B\rangle\langle C,D\rangle \nonumber\\
&&\nonumber\\
&&\qquad \, \, \, \, \, \, \,  \equiv {\cal{H}}(A,B,C,D) \, \langle A,B\rangle\langle C,D\rangle,
\end{eqnarray}
where
$$
\delta u = u_1 - u_2.
$$
\subsection{The Factor $R_E$}
We will now compute $R_E$. One has
\begin{eqnarray}\label{eqn;RF2}
\prod^{\rightarrow}_{n\geq0} \exp[-e_n\otimes f_{-1-n}]|k,l\rangle &=& \sum_{m} B_m |k+m,l-m\rangle.
\end{eqnarray}
Let us define as in the main text
\begin{eqnarray}
\tilde{c}_i &=& u_2-\frac{C-D+1}{2} -i,\\
\tilde{d}_i &=& u_1-\frac{A-B-1}{2} +i.
\end{eqnarray}
The term $B_m$ is this time built up out of $m$ copies of
$-e\otimes f$ acting on the state $\langle A,B\rangle\langle
C,D\rangle$. One has
\begin{eqnarray}
e_{n_m}\ldots e_{n_2} e_{n_1}\langle A,B\rangle &=&e_{n_m}\ldots e\left(u+\frac{h+1}{2}\right)^{n_2}e\left(u+\frac{h+1}{2}\right)^{n_1}\langle A,B\rangle \nonumber\\
&=& e_{n_m}\ldots e\left(u+\frac{h+1}{2}\right)^{n_2} e \left({\tilde{d}}_0\right)^{n_1}\langle A,B\rangle \nonumber\\
&=& A\left(\tilde{d}_0\right)^{n_1}e_{n_m}\ldots e\left(u+\frac{h+1}{2}\right)^{n_2}\langle A-1,B+1\rangle \nonumber\\
&=& A(A-1)\left(\tilde{d}_0\right)^{n_1} \left(\tilde{d}_1\right)^{n_2}e_{n_m}\ldots e_{n_3}\langle A-2,B+2\rangle \nonumber\\
&=& \frac{A!}{(A-m)!} \tilde{d}_0^{n_1}\ldots \tilde{d}_{m-1}^{n_m}\langle A-m,B+m\rangle.
\end{eqnarray}
Similar expressions hold for $f_n$ acting on $\langle C,D\rangle$,
with $\tilde{c}_i$ instead of $\tilde{d}_i$, and producing the
state $\langle C+m,D-m\rangle$. From the ordered exponential
(\ref{eqn;RF2}) we have now $n_i\leq n_{i-1}$. In case $n_i =
n_{i+1}$, we again pick up the same combinatorial factor as in the calculation of $R_F$, coming
from the series of the exponential. Putting all of this together,
we find
\begin{eqnarray}
&&B_m =  \frac{A!}{(A-m)!}\frac{D!}{(D-m)!}\left\{ \sum_{n_1\geq\ldots\geq n_m} \frac{1}{N(\{n_1,\ldots,n_m\})}\frac{\tilde{d}_0^{n_1}}{\tilde{c}_0^{{}\, n_1+1}}\ldots \frac{\tilde{d}_{m-1}^{n_m}}{\tilde{c}_{m-1}^{{}\, n_m+1}} \right\}, \nonumber\\
&&N(\{n_1,\ldots,n_m\})=\frac{1}{{\rm ord} S(\{n_1,\ldots,n_m\})},
\end{eqnarray}
where $N$ is defined as in the formulas for $R_F$. The sum
evaluates at
\begin{eqnarray}
B_m (A,B,C,D) = m! {A\choose m}{D\choose m}
\prod_{i=0}^{m-1}\frac{1}{\tilde{d}_0-\tilde{c}_0-i+m-1}.
\end{eqnarray}

\bibliographystyle{JHEP}
\bibliography{LitRmat}

\providecommand{\href}[2]{#2}\begingroup\raggedright\begin{thebibliography}{10}

\bibitem{Minahan:2002ve}
J.~A. Minahan and K.~Zarembo, {\it The \textrm{B}ethe-ansatz for $\mathcal{N} =
  4$ super \textrm{Y}ang-\textrm{M}ills},  {\em JHEP} {\bf 03} (2003) 013,
  [\href{http://xxx.lanl.gov/abs/hep-th/0212208}{{\tt hep-th/0212208}}].

\bibitem{Kazakov:2004qf}
V.~A. Kazakov, A.~Marshakov, J.~A. Minahan, and K.~Zarembo, {\it Classical /
  quantum integrability in $\mathit{AdS/CFT}$},  {\em JHEP} {\bf 05} (2004)
  024, [\href{http://xxx.lanl.gov/abs/hep-th/0402207}{{\tt hep-th/0402207}}].

\bibitem{Beisert:2004hm}
N.~Beisert, V.~Dippel, and M.~Staudacher, {\it A novel long range spin chain
  and planar $\mathcal{N} = 4$ super \textrm{Y}ang- \textrm{M}ills},  {\em
  JHEP} {\bf 07} (2004) 075,
  [\href{http://xxx.lanl.gov/abs/hep-th/0405001}{{\tt hep-th/0405001}}].

\bibitem{Arutyunov:2004vx}
G.~Arutyunov, S.~Frolov, and M.~Staudacher, {\it Bethe ansatz for quantum
  strings},  {\em JHEP} {\bf 10} (2004) 016,
  [\href{http://xxx.lanl.gov/abs/hep-th/0406256}{{\tt hep-th/0406256}}].

\bibitem{Staudacher:2004tk}
M.~Staudacher, {\it The factorized $\mathit{S}$-matrix of \textrm{CFT/AdS}},
  {\em JHEP} {\bf 05} (2005) 054,
  [\href{http://xxx.lanl.gov/abs/hep-th/0412188}{{\tt hep-th/0412188}}].

\bibitem{Beisert:2005fw}
N.~Beisert and M.~Staudacher, {\it Long-range $\mathfrak{psu}(2,2|4)$
  \textrm{B}ethe ansaetze for gauge theory and strings},  {\em Nucl. Phys.}
  {\bf B727} (2005) 1--62, [\href{http://xxx.lanl.gov/abs/hep-th/0504190}{{\tt
  hep-th/0504190}}].

\bibitem{Beisert:2005tm}
N.~Beisert, {\it {The $su(2|2)$ dynamic S-matrix}},  {\em Adv. Theor. Math.
  Phys.} {\bf 12} (2008) 945,
  [\href{http://xxx.lanl.gov/abs/hep-th/0511082}{{\tt hep-th/0511082}}].

\bibitem{Hofman:2006xt}
D.~M. Hofman and J.~M. Maldacena, {\it {Giant magnons}},  {\em J. Phys.} {\bf
  A39} (2006) 13095--13118, [\href{http://xxx.lanl.gov/abs/hep-th/0604135}{{\tt
  hep-th/0604135}}].

\bibitem{Arutyunov:2006ak}
G.~Arutyunov, S.~Frolov, J.~Plefka, and M.~Zamaklar, {\it The off-shell
  symmetry algebra of the light-cone $\mathit{AdS}_{5}\times \mathit{S}^5$
  superstring},  {\em J. Phys.} {\bf A40} (2007) 3583--3606,
  [\href{http://xxx.lanl.gov/abs/hep-th/0609157}{{\tt hep-th/0609157}}].

\bibitem{Klose:2006zd}
T.~Klose, T.~McLoughlin, R.~Roiban, and K.~Zarembo, {\it Worldsheet scattering
  in $\mathit{AdS}_{5}\times \mathit{S}^5$},  {\em JHEP} {\bf 03} (2007) 094,
  [\href{http://xxx.lanl.gov/abs/hep-th/0611169}{{\tt hep-th/0611169}}].

\bibitem{Arutyunov:2009ga}
G.~Arutyunov and S.~Frolov, {\it {Foundations of the $\mathit{AdS}_{5}\times
  \mathit{S}^5$ Superstring. Part I}},
  \href{http://xxx.lanl.gov/abs/0901.4937}{{\tt arXiv:0901.4937}}.

\bibitem{Janik:2006dc}
R.~A. Janik, {\it The $\mathit{AdS}_{5}\times \mathit{S}^5$ superstring
  worldsheet $\mathit{S}$-matrix and crossing symmetry},  {\em Phys. Rev.} {\bf
  D73} (2006) 086006, [\href{http://xxx.lanl.gov/abs/hep-th/0603038}{{\tt
  hep-th/0603038}}].

\bibitem{Beisert:2006ib}
N.~Beisert, R.~Hernandez, and E.~Lopez, {\it A crossing-symmetric phase for
  $\mathit{AdS}_{5}\times \mathit{S}^5$ strings},  {\em JHEP} {\bf 11} (2006)
  070, [\href{http://xxx.lanl.gov/abs/hep-th/0609044}{{\tt hep-th/0609044}}].

\bibitem{Beisert:2006ez}
N.~Beisert, B.~Eden, and M.~Staudacher, {\it Transcendentality and crossing},
  {\em J. Stat. Mech.} {\bf 0701} (2007) P021,
  [\href{http://xxx.lanl.gov/abs/hep-th/0610251}{{\tt hep-th/0610251}}].

\bibitem{Bajnok:2008bm}
Z.~Bajnok and R.~A. Janik, {\it {Four-loop perturbative Konishi from strings
  and finite size effects for multiparticle states}},  {\em Nucl. Phys.} {\bf
  B807} (2009) 625--650, [\href{http://xxx.lanl.gov/abs/0807.0399}{{\tt
  arXiv:0807.0399}}].

\bibitem{Fiamberti:2007rj}
F.~Fiamberti, A.~Santambrogio, C.~Sieg, and D.~Zanon, {\it {Wrapping at four
  loops in N=4 SYM}},  {\em Phys. Lett.} {\bf B666} (2008) 100--105,
  [\href{http://xxx.lanl.gov/abs/0712.3522}{{\tt arXiv:0712.3522}}].

\bibitem{Zwiebel:2008gr}
B.~I. Zwiebel, {\it {Iterative Structure of the N=4 SYM Spin Chain}},  {\em
  JHEP} {\bf 07} (2008) 114, [\href{http://xxx.lanl.gov/abs/0806.1786}{{\tt
  arXiv:0806.1786}}].

\bibitem{Bajnok:2008qj}
Z.~Bajnok, R.~A. Janik, and T.~Lukowski, {\it {Four loop twist two, BFKL,
  wrapping and strings}},  \href{http://xxx.lanl.gov/abs/0811.4448}{{\tt
  arXiv:0811.4448}}.

\bibitem{Arutyunov:2009zu}
G.~Arutyunov and S.~Frolov, {\it {String hypothesis for the
  $\mathit{AdS}_{5}\times \mathit{S}^5$ mirror}},
  \href{http://xxx.lanl.gov/abs/0901.1417}{{\tt arXiv:0901.1417}}.

\bibitem{Gromov:2009tv}
N.~Gromov, V.~Kazakov, and P.~Vieira, {\it {Integrability for the Full Spectrum
  of Planar AdS/CFT}},  \href{http://xxx.lanl.gov/abs/0901.3753}{{\tt
  arXiv:0901.3753}}.

\bibitem{Bargheer:2009xy}
T.~Bargheer, N.~Beisert, and F.~Loebbert, {\it {Long-Range Deformations for
  Integrable Spin Chains}},  \href{http://xxx.lanl.gov/abs/0902.0956}{{\tt
  arXiv:0902.0956}}.

\bibitem{Bombardelli:2009ns}
D.~Bombardelli, D.~Fioravanti, and R.~Tateo, {\it {Thermodynamic Bethe Ansatz
  for planar AdS/CFT: a proposal}},
  \href{http://xxx.lanl.gov/abs/0902.3930}{{\tt arXiv:0902.3930}}.

\bibitem{Gromov:2009bc}
N.~Gromov, V.~Kazakov, A.~Kozak, and P.~Vieira, {\it {Integrability for the
  Full Spectrum of Planar AdS/CFT II}},
  \href{http://xxx.lanl.gov/abs/0902.4458}{{\tt arXiv:0902.4458}}.

\bibitem{Arutyunov:2009ur}
G.~Arutyunov and S.~Frolov, {\it {Thermodynamic Bethe Ansatz for the
  $\mathit{AdS}_{5}\times \mathit{S}^5$ Mirror Model}},
  \href{http://xxx.lanl.gov/abs/0903.0141}{{\tt arXiv:0903.0141}}.

\bibitem{Arutyunov:2009mi}
G.~Arutyunov, M.~de~Leeuw, and A.~Torrielli, {\it {The Bound State S-Matrix for
  $\mathit{AdS}_{5}\times \mathit{S}^5$ Superstring}},
  \href{http://xxx.lanl.gov/abs/0902.0183}{{\tt arXiv:0902.0183}}.

\bibitem{Dorey:2006dq}
N.~Dorey, {\it {Magnon bound states and the AdS/CFT correspondence}},  {\em J.
  Phys.} {\bf A39} (2006) 13119--13128,
  [\href{http://xxx.lanl.gov/abs/hep-th/0604175}{{\tt hep-th/0604175}}].

\bibitem{Drummond:2009fd}
J.~M. Drummond, J.~M. Henn, and J.~Plefka, {\it {Yangian symmetry of scattering
  amplitudes in N=4 super Yang-Mills theory}},
  \href{http://xxx.lanl.gov/abs/0902.2987}{{\tt arXiv:0902.2987}}.

\bibitem{Torrielli:2007mc}
A.~Torrielli, {\it {Classical r-matrix of the $su(2|2)$ SYM spin-chain}},  {\em
  Phys. Rev.} {\bf D75} (2007) 105020,
  [\href{http://xxx.lanl.gov/abs/hep-th/0701281}{{\tt hep-th/0701281}}].

\bibitem{Beisert:2007ds}
N.~Beisert, {\it {The S-Matrix of AdS/CFT and Yangian Symmetry}},  {\em PoS}
  {\bf SOLVAY} (2006) 002, [\href{http://xxx.lanl.gov/abs/0704.0400}{{\tt
  0704.0400}}].

\bibitem{deLeeuw:2008dp}
M.~de~Leeuw, {\it {Bound States, Yangian Symmetry and Classical r-matrix for
  the $\mathit{AdS}_{5}\times \mathit{S}^5$ Superstring}},  {\em JHEP} {\bf 06}
  (2008) 085, [\href{http://xxx.lanl.gov/abs/0804.1047}{{\tt
  arXiv:0804.1047}}].

\bibitem{Moriyama:2007jt}
S.~Moriyama and A.~Torrielli, {\it {A Yangian Double for the AdS/CFT Classical
  r-matrix}},  {\em JHEP} {\bf 06} (2007) 083,
  [\href{http://xxx.lanl.gov/abs/0706.0884}{{\tt 0706.0884}}].

\bibitem{Matsumoto:2007rh}
T.~Matsumoto, S.~Moriyama, and A.~Torrielli, {\it {A Secret Symmetry of the
  AdS/CFT S-matrix}},  {\em JHEP} {\bf 09} (2007) 099,
  [\href{http://xxx.lanl.gov/abs/0708.1285}{{\tt arXiv:0708.1285}}].

\bibitem{Beisert:2007ty}
N.~Beisert and F.~Spill, {\it {The Classical r-matrix of AdS/CFT and its Lie
  Bialgebra Structure}},  {\em Commun. Math. Phys.} {\bf 285} (2009) 537--565,
  [\href{http://xxx.lanl.gov/abs/0708.1762}{{\tt arXiv:0708.1762}}].

\bibitem{Spill:2008tp}
F.~Spill and A.~Torrielli, {\it {On Drinfeld's second realization of the
  AdS/CFT $su(2|2)$ Yangian}},  {\em J. Geom. Phys. (in press)}
  [\href{http://xxx.lanl.gov/abs/0803.3194}{{\tt arXiv:0803.3194}}].

\bibitem{Matsumoto:2008ww}
T.~Matsumoto and S.~Moriyama, {\it {An Exceptional Algebraic Origin of the
  AdS/CFT Yangian Symmetry}},  {\em JHEP} {\bf 04} (2008) 022,
  [\href{http://xxx.lanl.gov/abs/0803.1212}{{\tt arXiv:0803.1212}}].

\bibitem{deLeeuw:2008ye}
M.~de~Leeuw, {\it {The Bethe Ansatz for $\mathit{AdS}_{5}\times \mathit{S}^5$
  Bound States}},  {\em JHEP} {\bf 01} (2009) 005,
  [\href{http://xxx.lanl.gov/abs/0809.0783}{{\tt arXiv:0809.0783}}].

\bibitem{Spill:2008yr}
F.~Spill, {\it {Weakly coupled N=4 Super Yang-Mills and N=6 Chern-Simons
  theories from $u(2|2)$ Yangian symmetry}},
  \href{http://xxx.lanl.gov/abs/0810.3897}{{\tt arXiv:0810.3897}}.

\bibitem{Matsumoto:2009rf}
T.~Matsumoto and S.~Moriyama, {\it {Serre Relation and Higher Grade Generators
  of the AdS/CFT Yangian Symmetry}},
  \href{http://xxx.lanl.gov/abs/0902.3299}{{\tt arXiv:0902.3299}}.

\bibitem{Arutyunov:2007tc}
G.~Arutyunov and S.~Frolov, {\it {On String S-matrix, Bound States and TBA}},
  {\em JHEP} {\bf 12} (2007) 024,
  [\href{http://xxx.lanl.gov/abs/0710.1568}{{\tt 0710.1568}}].

\bibitem{twi}
V.~G. Drinfeld, {\it {Quasi-Hopf algebras}},  {\em Leningrad Math. J.} {\bf 1}
  (1990) 1419.

\bibitem{Khoroshkin:1994uk}
S.~M. Khoroshkin and V.~N. Tolstoy, {\it {Yangian double and rational R
  matrix}},  \href{http://xxx.lanl.gov/abs/hep-th/9406194}{{\tt
  hep-th/9406194}}.

\bibitem{Arutyunov:2008zt}
G.~Arutyunov and S.~Frolov, {\it {The S-matrix of String Bound States}},  {\em
  Nucl. Phys.} {\bf B804} (2008) 90--143,
  [\href{http://xxx.lanl.gov/abs/0803.4323}{{\tt arXiv:0803.4323}}].

\bibitem{Kulish:1981gi}
P.~P. Kulish, N.~Y. Reshetikhin, and E.~K. Sklyanin, {\it {Yang-Baxter Equation
  and Representation Theory. 1}},  {\em Lett. Math. Phys.} {\bf 5} (1981)
  393--403.

\bibitem{Chari}
V.~Chari and A.~Pressley, {\it {A Guide To Quantum Groups}},  {\em Cambridge,
  UK: Univ. Press} (1994).

\bibitem{Chen:2006gq}
H.-Y. Chen, N.~Dorey, and K.~Okamura, {\it {On the scattering of magnon
  boundstates}},  {\em JHEP} {\bf 11} (2006) 035,
  [\href{http://xxx.lanl.gov/abs/hep-th/0608047}{{\tt hep-th/0608047}}].

\bibitem{Gomez:2006va}
C.~Gomez and R.~Hernandez, {\it {The magnon kinematics of the AdS/CFT
  correspondence}},  {\em JHEP} {\bf 11} (2006) 021,
  [\href{http://xxx.lanl.gov/abs/hep-th/0608029}{{\tt hep-th/0608029}}].

\bibitem{Plefka:2006ze}
J.~Plefka, F.~Spill, and A.~Torrielli, {\it {On the Hopf algebra structure of
  the AdS/CFT S-matrix}},  {\em Phys. Rev.} {\bf D74} (2006) 066008,
  [\href{http://xxx.lanl.gov/abs/hep-th/0608038}{{\tt hep-th/0608038}}].

\bibitem{Arutyunov:2006yd}
G.~Arutyunov, S.~Frolov, and M.~Zamaklar, {\it {The Zamolodchikov-Faddeev
  algebra for $\mathit{AdS}_{5}\times \mathit{S}^5$ superstring}},  {\em JHEP}
  {\bf 04} (2007) 002, [\href{http://xxx.lanl.gov/abs/hep-th/0612229}{{\tt
  hep-th/0612229}}].

\bibitem{Beisert:2005wm}
N.~Beisert, {\it {An $SU(1|1)$-invariant S-matrix with dynamic
  representations}},  {\em Bulg. J. Phys.} {\bf 33S1} (2006) 371--381,
  [\href{http://xxx.lanl.gov/abs/hep-th/0511013}{{\tt hep-th/0511013}}].

\bibitem{stuko}
V.~Stukopin, {\it {Yangians of classical lie superalgebras: Basic
  constructions, quantum double and universal R-matrix}},  {\em Proceedings of
  the Institute of Mathematics of NAS of Ukraine} {\bf 50} (2004) 1195.

\bibitem{Heckenberger:2007ry}
I.~Heckenberger, F.~Spill, A.~Torrielli, and H.~Yamane, {\it {Drinfeld second
  realization of the quantum affine superalgebras of $D^{(1)}(2,1:x)$ via the
  Weyl groupoid}},  {\em RIMS Kokyuroku Bessatsu} {\bf B8} (2008) 171--216,
  [\href{http://xxx.lanl.gov/abs/0705.1071}{{\tt arXiv:0705.1071}}].

\bibitem{Gow}
L.~Gow, {\it {Gauss Decomposition of the Yangian $Y(\mathfrak{gl}(m|n))$}},
  {\em Commun. Math. Phys.} {\bf 276} (2007) 799--825.

\bibitem{Etingof}
P.~Etingof and O.~Schiffman, {\it {Lectures on Quantum Groups}},  {\em Lectures
  in Mathematical Physics, International Press, Boston} (1998).

\bibitem{MacKay:2004tc}
N.~J. MacKay, {\it {Introduction to Yangian symmetry in integrable field
  theory}},  {\em Int. J. Mod. Phys.} {\bf A20} (2005) 7189--7218,
  [\href{http://xxx.lanl.gov/abs/hep-th/0409183}{{\tt hep-th/0409183}}].

\bibitem{Molev}
A.~Molev, {\it {Yangians and Classical Lie Algebras}},  {\em Mathematical
  Surveys and Monographs 143, American Mathematical Society, Providence, RI}
  (2007).

\bibitem{Drin}
V.~G. Drinfeld, {\it {Quantum groups}},  {\em Proc. of the International
  Congress of Mathematicians, Berkeley, 1986, American Mathematical Society}
  (1987) 798.

\bibitem{Dsecond}
V.~G. Drinfeld, {\it {A new realization of Yangians and quantum affine
  algebras}},  {\em Soviet Math. Dokl.} {\bf 36} (1988) 212.

\bibitem{KT}
S.~M. Khoroshkin and V.~N. Tolstoy, {\it {Universal R-matrix for quantized
  (super)algebras}},  {\em Commun. Math. Phys.} {\bf 276} (1991) 599--617.

\bibitem{Cai:q-alg9709038}
J.~Cai, S.~Wang, K.~Wu, and C.~Xiong, {\it {Universal R-matrix Of The Super
  Yangian Double $DY(gl(1|1))$}},  {\em Comm. Theor. Phys.} {\bf 29} (1998)
  173--176, [\href{http://xxx.lanl.gov/abs/q-alg/9709038}{{\tt
  q-alg/9709038}}].

\bibitem{Torrielli:2008wi}
A.~Torrielli, {\it {Structure of the string R-matrix}},  {\em J. Phys.} {\bf
  A42} (2009) 055204, [\href{http://xxx.lanl.gov/abs/0806.1299}{{\tt
  arXiv:0806.1299}}].

\end{thebibliography}\endgroup

\end{document}